\documentclass[a4paper,12pt]{article}

\usepackage{graphicx}
\usepackage[T1]{fontenc}
\usepackage[english]{babel}
\usepackage{hyperref}
\def\a{\alpha}

\def\be{\begin{equation}}            \def\ee{\end{equation}}
\def\ba#1{\begin{array}{#1}}         \def\ea{\end{array}}

\begin{document}
\noindent
\centerline{\Large\bf
TRI-CRYSTAL FIXED EXIT MONOCHROMATOR
}\\[2.5mm]
\centerline{I.L. Zhogin
 }\\[2.2mm]
\centerline{\em Institute of Solid State Chemistry
and
Mechanochemistry SB RAS, Novosibirsk, Russia }
\thispagestyle{empty}
\noindent

It is proposed a novel design of an X-ray monochromator  which
uses three crystals marked with $A,B,C$ on the figure 1a. The
exit beam, $CD$, turns in horizontal plane on the angle $2\a$
with respect to the incoming beam, $SA$ (and the segment $AC$
bisects this angle, see the fig.~1a). So, this kind of
monochromator is well suitable for side beam-lines;
several tunable beam-lines can be built for a one fan beam
of synchrotron radiation (from wiggler  or bending magnet).

The first and the last crystals are equal (of the same
interplanar spacing $d$), and fixed,
while the second one is moving in the middle plane
($|AB|=|BC|$)
  and has about half
spacing, $d^*\approx d/2$. The next sets of crystals are well
appropriate:
\[ \mbox{Si-111, Si-311, Si-111, in this case }
k^2=(d^*/d)^2=3/11;\]
\[ 2\times\mbox{Si-220$\,+\,$Si-440}, \
k^2=0.25;\]
\[ 2\times \mbox{Si-311$\,+\,$Si-533
(or Si-444)}, \ k^2=11/43 \ ({\rm or\ }11/48).
\]

The tilt from vertical, angle $\varphi$, and the length
$h=|OB|$  define completely the second crystal's position.
It is convenient to take the unit of length as follows:
\[
|AO|=|AC|/2=1;\]
 this unit can be as small
as the crystal's sizes, that is about 4--5 cm.\\[-5mm]
\begin{figure}[h]  
\begin{center}
\includegraphics[bb=116 80 338 182,width=100mm,height=55mm]{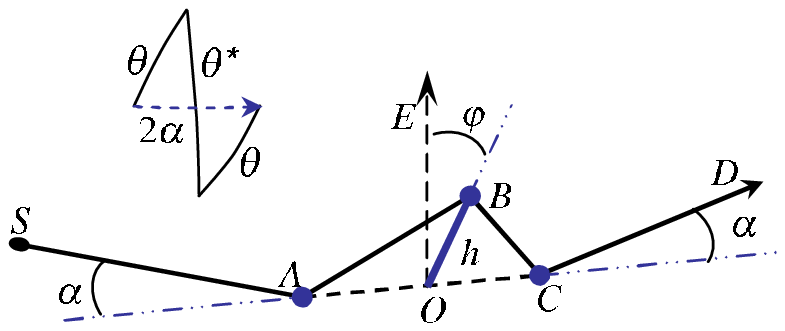}
\end{center}  \label{fig1}
\end{figure}

\vspace{-55mm}
\hspace*{25mm} (b)

\vspace{31mm}\hspace{11mm} (a)\\[-1mm]


\noindent {\bf Figure\,1.} The scheme of tri-crystal monochromator (a);
Bragg angles of the crystals, $\theta,\theta^*$,
 and the resulting turn of outgoing beam,
$\a=\,$const (b).\\[1.5mm]

Taking the point $O$ as the center of Cartesian coordinate system
(axis 3 is vertical, along $\overrightarrow{OE}$;
axis 1 goes along $\overrightarrow{AC}$),
one can write the crystals' coordinates:
\[ A=(-1,0,0),\ C=(1,0,0),\ B=(0,-h\sin\varphi,h\cos\varphi).\]
Then, the unit vectors along the beam after the first
and the second
reflection, $\vec{n}_1$ and $\vec{n}_2$, look as follows:
\[ \vec{n}_1
=\frac{(1,-h\sin\varphi,h\cos\varphi)}{\sqrt{1+h^2}}
\propto \overrightarrow{AB\,},\
\vec{n}_2
=\frac{(1,h\sin\varphi,-h\cos\varphi)}{\sqrt{1+h^2}}
\propto \overrightarrow{BC\,}, \]
while the unit vectors along incoming and outgoing beams read:
\[ \vec{n}_{\rm in} = (\cos\a,-\sin\a,0),
\ \vec{n}_{\rm out} = (\cos\a,\sin\a,0),
\]
The Bragg angles of crystals, $\theta$ and $\theta^*$
(the odd crystals have the same Bragg
 angle, $\theta$, because of symmetry
-- mirror plane permutes $A$ and $C$) can be find
straightforwardly:
\be \label{theta}
\cos2\theta=(\vec{n}_{\rm in}\cdot\vec{n}_1)=
\frac{\cos\a+h\sin\varphi\,\sin\a}{\sqrt{1+h^2}},
\ee
\be \label{theta*}
\cos2\theta^*=(\vec{n}_{1}\cdot\vec{n}_2)=
\frac{1-h^2}{1+h^2}=1-2\sin^2\!\theta^*,
\ee
These angles should be (energy)
related by means of the next requirement:
\be
d\sin\theta=d^*\sin\theta^*, \mbox{ \ or }
k= \frac{\,d^*}d=\frac{\,\sin\theta}{\sin\theta^*}\,.
\ee
The figures~2 and 3 shows $h$ and $\varphi$
as functions of the Bragg angle $\theta$
for two choices of $k$;
 these functions follow from Eqs.~(1)--(3) and
 read as follows:
 \be  
\ \ \ \ h=\frac{\sin \theta}{\sqrt{k^2-\sin^2\!\theta}}\,, \
\ \sin \varphi= \frac{\sqrt{1+h^2}\cos2\theta - \cos\a}
   {h\sin\a}\,.
\ee
 \begin{figure}[h]  
  \begin{center}
  \includegraphics[bb=65 33 530 365,width=95mm]{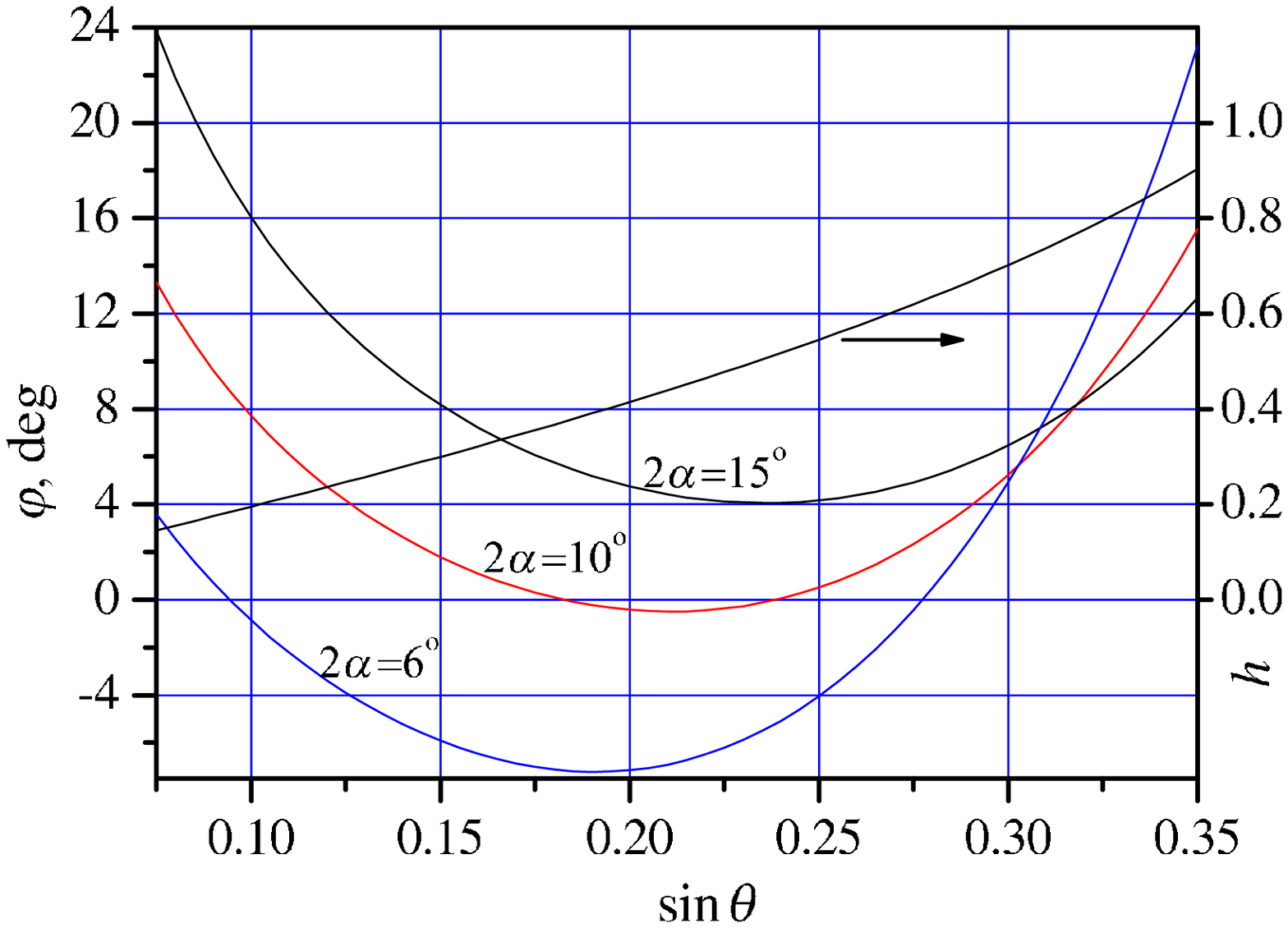}
 \end{center}
  \label{fig2}
\end{figure}\\[-4mm]
\noindent {\bf Figure\,2.} The plot
$h(\sin \theta)$, and the curves
$\varphi(\sin \theta;\,\a)$ for different $\a$;
$k^2=3/11$.\\[0.5mm]

It is obvious that very small travel of the second crystal, 1--2$\,$cm,
suffices for doubling $\sin\theta$ (and energy of the exit beam);
so, one can use very small stage driven by piezo motor.
A usual double crystal fixed exit monochromator needs much more
lengthy positioner, see e.g. [1].

The tilt from  vertical of the reflection plane
of the odd crystals, the angle
$\psi$, also can be find from considering the height of
the point $B$:
\[H=h\cos\varphi=\sqrt{1+h^2}\,\sin2\theta\,\cos\psi, \
\cos\psi=\frac{\cos\varphi}{2k\cos\theta}.\]
The more robust expression gives also the sign
of this angle (the positive direction is chosen
inward, i.e. toward the letter $E$):
\[ \sin\psi=\frac{\sin\a-h\,\sin\varphi\,
\cos\a}{2kh\cos\theta}\,.
\]

 \begin{figure}[h]  
  \begin{center}
  \includegraphics[bb=60 33 532 365,width=95mm]{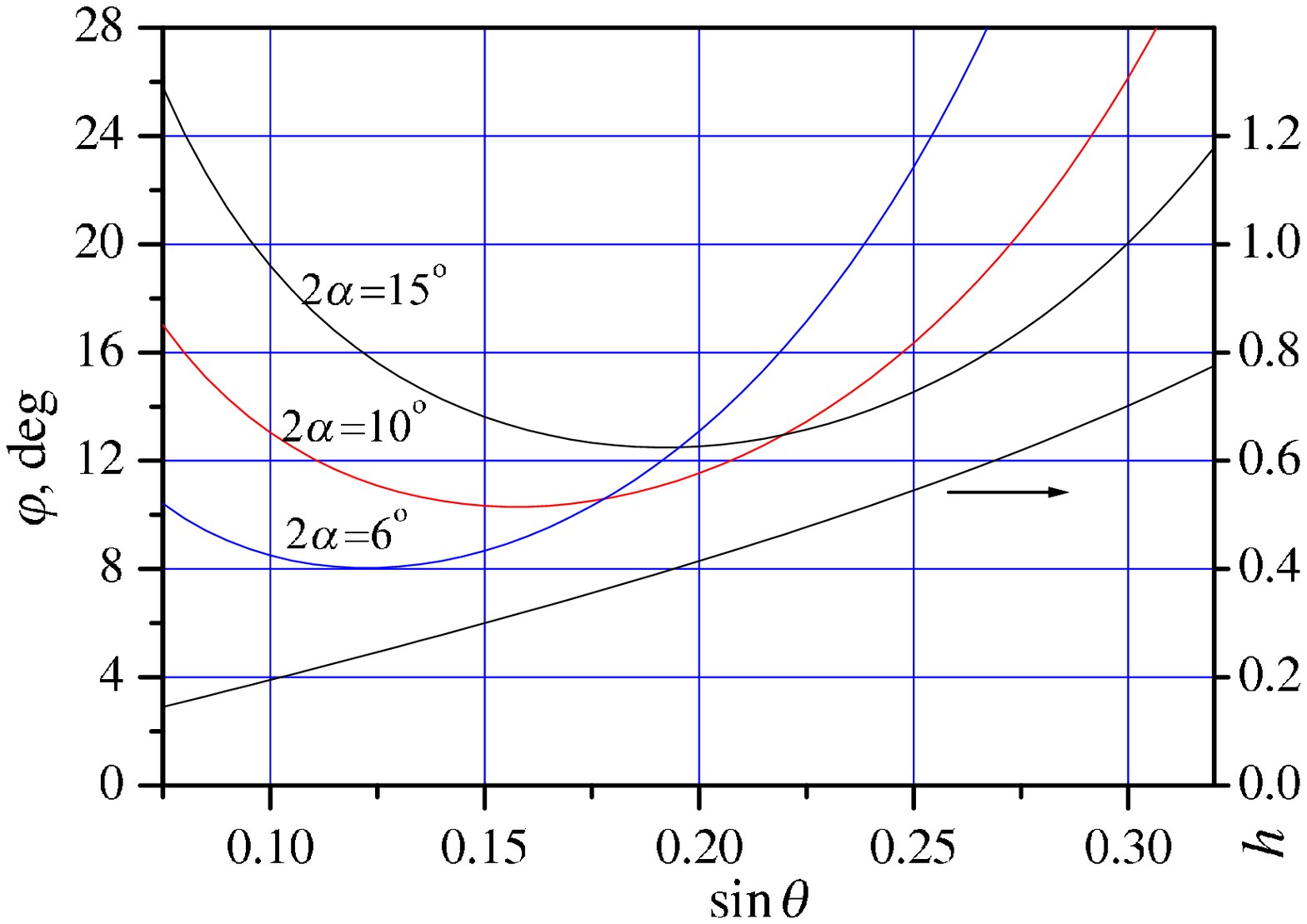}
 \end{center}
  \label{fig3}
\end{figure}

\vspace{-4mm}

\noindent {\bf Figure\,3.} The plot
$h(\sin \theta)$, and the curves
$\varphi(\sin \theta;\,\a)$ for different $\a$;
$k^2=0.25$.\\[0.5mm]

In contrast to the double-crystal scheme of general position considered
in [2] (which also suitable for
side beam-lines),
the new variant does not lead to a tilt of exit beam profile
(and its polarization): it is clear from fig.~1b that two spherical
triangles formed by arcs $\a^{-1}\!,\theta,\theta^*\!,\theta$ cancel one
another
(they are path-traced in opposite directions).

One may think also
about bent crystals as a way to focus  the exit beam; this option is
especially important for hard X-ray range where grazing mirrors do not work.
\\[2mm]
\rule{5cm}{0.2mm}\\[2mm]

\noindent
[1] N. Gavrilov, I. Zhogin, A. Shmakov et al.,
"Lay-out of ultrahigh-vacuum DC-monochro\-mator",
SRI-03 proc., ed. by T. Warwick et al. AIP Conf. Proc.,
V.705. Melville, NY-2004, 691--694.

\noindent
[2] N.\,Gavrilov, I.\,Zhogin, M.\,Sheromov, B.\,Tolochko,
Nucl.\,Instr.\ and Meth., {\bf A543} (2005) 375-380;
\href{http://www.arXiv.org/abs/physics/0306191}%
{arXiv.org/physics/0306191}.


\end{document}